\documentclass[seceq]{ptptex}

\usepackage{graphicx}

\newcommand{\nc}{\newcommand}		
\nc{\vc}[1]	{\mbox{\boldmath $#1$}}	

\notypesetlogo                       

\title{
Level Density in the Complex Scaling Method
}

\author{
Ryusuke \textsc{Suzuki},$^{1}$
Takayuki \textsc{Myo}$^{2}$ and
Kiyoshi \textsc{Kat\=o}$^{1}$
}    
\inst{
$^{1}$Division of Physics, Graduate School of Science, Hokkaido University,\\
Sapporo 060-0810, Japan\\
$^{2}$Research Center for Nuclear Physics (RCNP), Osaka University,
Ibaraki 567-0047, Japan 
}   

\recdate{February 3, 2005}
\abst{
It is shown that the continuum level density (CLD) at unbound energies
can be calculated with the complex scaling method (CSM), in which the
energy spectra of bound states, resonances and continuum states are
obtained in terms of $L^2$ basis functions.
In this method, the extended completeness relation is applied to the
calculation of the Green functions, and the continuum-state part is
approximately expressed in terms of discretized complex scaled continuum
solutions.
The obtained result is compared with the CLD calculated exactly from the
scattering phase shift.
The discretization in the CSM is shown to give a very good description
of continuum states.
We discuss how the scattering phase shifts can inversely be calculated
from the discretized CLD using a basis function technique in the CSM.
}

\begin{document}     

\maketitle 

\section{Introduction}
Recently, there has been much interest in nuclear structures of unstable
nuclei, in which exotic nuclear structures have been revealed through the
development of radioactive nuclear beam experiments.\cite{Ta96}
It has been shown that for such nuclei, for example the so-called
neutron halo nuclei, there are extremely weak binding ground states, and
most of the excited states are in the continuum energy region.
Therefore, to understand the exotic structures and excitations of these
nuclei, it is necessary to study continuum and resonant states in
unbound energy regions. 
 
The continuum level density (CLD) is expected to play an important role
in relating experimental data and theoretical models for unbound states.
Recently, Kruppa and Arai\cite{Kr98,Kr99,Ar99} proposed an interesting
method to calculate the CLD and argued that resonance parameters can be
determined from the CLD.
They start their investigation from the definition of the CLD,\cite{Sh92}
\begin{equation}
 \Delta(E)
  =
  -\frac{1}{\pi}{\rm Im}
  \left[{\rm Tr}\left[G(E)-G_0(E)\right]\right],
 \label{delta1}
\end{equation}
where the full and free Green functions are given by $G(E)=(E-H)^{-1}$
and $G_0(E)=(E-H_0)^{-1}$, respectively.
Because the Hamiltonian $H$ includes finite range interactions in
addition to the asymptotic Hamiltonian $H_0$, the CLD expresses the
effect from the interactions.
When the eigenvalues ($\epsilon_i$ and $\epsilon_0^j$, respectively) of
$H$ and $H_0$ are obtained approximately within a framework including a
finite number ($N$) of basis functions, the following discrete level
density is defined:
\begin{equation}
 \Delta_N(E)
  =
  \sum_i^N\delta(E-\epsilon_i)-\sum_j^N\delta(E-\epsilon_0^j).
  \label{delta2}
\end{equation}
Kruppa\cite{Kr98} employed a smoothing technique defined by the
Strutinsky procedure\cite{St67} to calculate the continuous CLD,
$\Delta(E)$, from its discrete form, $\Delta_N(E)$.
However, as discussed in Ref.~\citen{Ar99} their results for the CLD
exhibit a strong dependence on the smoothing parameters.
We desire a more effective method to smooth the discrete quantities, or
to discretize the continuum states.

In this paper, we study a more direct method to calculate the CLD with
no smoothing technique in the framework of complex scaling,\cite{Ag71}
in which a basis function method is used to obtain not only bound
states but also resonance and continuum states.
The idea for the present method is taken from the extended completeness
relation,\cite{My98} originally proposed by Berggren,\cite{Be68} for
bound, resonance and continuum states in the complex scaling method
(CSM).
Exact proofs of this extended completeness relation for the CSM were
recently given for a coupled channel system\cite{Gi04} and a single
channel system\cite{Gi03}.
Green functions can be expressed by using the extended completeness
relation in terms of discrete eigenvalues of the CSM with a finite
number of basis functions.
Because the complex scaled Hamiltonians $H^\theta$ and $H_0^\theta$ have
complex eigenvalues, singularities, like the $\delta$-function contained
in Eq.~(\ref{delta2}), are avoided and replaced by Lorentzian functions.
Therefore, no smoothing process is needed.
Furthermore, it is shown that $\Delta(E)$ for the CLD can be calculated
independently of the scaling parameters in the CSM.

Kruppa and Arai\cite{Kr98,Kr99,Ar99} applied the CLD to search for
resonance parameters.
However, although parameter values for narrow resonances can be obtained
using any method of CSM and CLD, it is not easy to extract parameter
values for broad resonances with the CLD.
Rather than obtaining such resonance parameters, it is more important to
calculate the phase shift and/or S-matrix for the scattering states from
the CLD.
The CLD $\Delta(E)$ is related to the scattering phase shift $\delta(E)$
as\cite{Sh92,Le69}
\begin{eqnarray}
 \Delta(E) &=& \frac{1}{\pi}\frac{d\delta(E)}{dE}.
  \label{eq:exact}
\end{eqnarray}
Therefore, once we confirm that the CLD calculated in the present method
is consistent with $\Delta(E)$ obtained from the phase shift of the
scattering solution, we can inversely calculate the phase shift by
integrating the CLD obtained as a function of the energy from the
eigenvalues of the complex scaled Hamiltonians.
This implies that the phase shift can be calculated from discrete
eigenvalues using a basis function method.

We here demonstrate the reliability of this method by applying it to
several two-body systems, including $^4$He+$n$ and $\alpha$+$\alpha$,
which were studied by Arai and Kruppa.\cite{Ar99}
Comparing the calculated continuum level density and the phase shifts in
the CSM with the results obtained from the exactly calculated phase
shifts and their derivatives (the CLD), we show that the extended
completeness relation in the CSM is effective, and also that the
phase shift is satisfactorily reproduced by the discretized solutions
for continuum states.
  
In \S 2, we explain the formalism for treating the level density in the
complex scaling method.
We study the reliability of this method by applying it to a simple
potential model in \S 3, and we investigate $^4$He+$n$ and
$\alpha$+$\alpha$ systems in \S 4.
In \S 5, a summary and conclusions are given.

\section{Continuum level density in the complex scaling method}
Here we briefly explain the complex scaling method (CSM).
In the CSM, the spatial coordinate $\vc{r}$ and the wave number $\vc{k}$
transform as
\begin{eqnarray}
 U(\theta) : \vc{r} \rightarrow \vc{r}\exp(i\theta), \quad\quad
  \vc{k} \rightarrow \vc{k}\exp(-i\theta),
\end{eqnarray}
where $U(\theta)$ is a scaling operator and $\theta$ is a real number
called a scaling parameter.
Under this transformation, the asymptotic divergent behavior,
$\sim\exp(ik_rr)$, of a resonant state with a complex wave number
$k_r=\kappa-i\gamma$ is changed into a damping form,
$\exp{\{i(\kappa-i\gamma)(r\cos\theta+ir\sin\theta)\}}
=\exp{\{(\gamma\cos\theta-\kappa\sin\theta)r\}}
\cdot\exp{\{i(\kappa\cos\theta+\gamma\sin\theta)r\}}$
for $\theta>\tan^{-1}\gamma/\kappa$.
Therefore, resonant states and bound states are obtained as discrete
solutions of the complex scaled Schr\"odinger equation
\begin{eqnarray}
 H(\theta) \Phi^{\theta} = E\Phi^{\theta},
\end{eqnarray}
where $H(\theta)=U(\theta)H U^{-1}(\theta)$.
Because we require the complex scaled Hamiltonian to have no singularity, 
the scaling parameter $\theta$ has an upper limit, $\theta_{C}$.
For the Gaussian potential, $\theta_{C}=\pi/4$.
For $\theta<\theta_C$, the solutions of bound states and resonances with
$\gamma/\kappa<\tan\theta$ are square-integrable, because of their
damping forms in the asymptotic region.
Therefore, employing an appropriate scaling parameter $\theta$, we can
derive resonant states in addition to bound states using a
square-integrable basis expansion, for example, in terms of harmonic
oscillator or Gaussian functions $\{\phi_n\}$: 
\begin{equation}
 \Phi^{\theta}=\sum_{n=1}^Nc_n(\theta)\phi_n.
\end{equation}
\begin{figure}[tbp]
 \centerline{\includegraphics[width=9.0cm]{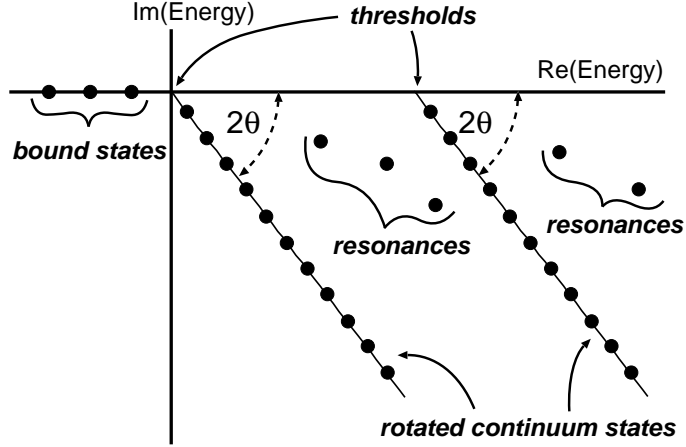}}
 \caption{
 Schematic energy eigenvalue distribution for a complex scaled
 Hamiltonian.
 }
 \label{fig:eigen}
\end{figure}
 
In Fig.~1, we present a schematic eigenvalue distribution for the
complex scaled Schr\"odinger equation. 
It is seen that the energies of bound states are not changed from the
spectral positions of the original Hamiltonian.  
The eigenvalues of the resonant states, which are particularly
noteworthy, are obtained as $E=E_{r}-i\Gamma_{r}/2$, where $E_{r}$ and
$\Gamma_{r}$ are the energy and width of a resonance, respectively.
By contrast, the continuum spectra of the Hamiltonian $H(\theta)$ are
distributed on the $2\theta$-lines originating from every threshold.
If we do not apply the complex scaling, the original Schr\"odinger
equation gives the continuum spectra, including resonances on the
positive energy axis.
Under complex scaling, the resonances for which
$\gamma/\kappa<\tan{\theta}$ are separated from the continuum, and the
rotated continuum spectra starting from different threshold energies are
separately obtained on different $2\theta$-lines.
Furthermore, when we apply a basis function method to solve the complex
scaled Schr\"odinger equation, these continuum spectra are discretized
on different $2\theta$-lines, as shown in Fig.~1.

Let us return to the problem of the level density.
The level density $\rho(E)$ of the Hamiltonian $H$ is defined as
\begin{equation}
 \rho(E)=\int\hspace{-0.5cm}\sum\delta(E-E_i),
\end{equation}
where the quantities $E_i$ are the eigenvalues of $H$, and the summation
and integration are taken for discrete and continuous eigenvalues,
respectively.
This definition of the level density can also be expressed using the
Green function:
\begin{eqnarray}
 \rho(E)
  &=&
  -\frac{1}{\pi}{\rm Im}
  \left\{ {\rm Tr}
   \left[\frac{1}{E-H}\right]\right\}
  \nonumber \\
 &=&
  -\frac{1}{\pi}{\rm Im} 
  \int d\vc{r}
  \left\langle\vc{r}\left|\frac{1}{E-H}\right|\vc{r}\right\rangle.
  \label{eq:lev_dens_CSM_1}
\end{eqnarray}
Here, applying the CSM and the extended completeness relation (ECR)
\cite{My98,Be68} to the expression of the Green function, we obtain
\begin{eqnarray}
 \rho(E)
  &=&
  -\frac{1}{\pi}{\rm Im} \int d\vc{r} 
  \left\langle\vc{r}
   \left|U(\theta)^{-1}U(\theta)\frac{1}{E-H}U(\theta)^{-1}U(\theta)\right|
   \vc{r}\right\rangle
  \nonumber\\
 &=& 
  -\frac{1}{\pi}{\rm Im} \int d\vc{r}
  \left\langle\vc{r}_\theta
   \left|\frac{1}{E-H(\theta)}\right|
   \vc{r}_\theta\right\rangle
  \nonumber\\
 &=&
  -\frac{1}{\pi}{\rm Im} \int d\vc{r} 
  \left[
   \sum_{B}^{N_B}
   \frac{\Phi^{\theta}_{B}(\vc{r}){\tilde \Phi}^{\theta*}_{B}(\vc{r})}{E-E_{B}}
   + 
   \sum_{R}^{N_{R}^{\theta}}
   \frac{\Phi^{\theta}_{R}(\vc{r}){\tilde \Phi}^{\theta*}_{R}(\vc{r})}{E-E_{R}}
   + 
   \int_{L_{\theta}}dk_{\theta}
   \frac{
   \Phi^{\theta}_{k_{\theta}}(\vc{r})
   {\tilde \Phi}^{\theta*}_{k_{\theta}}(\vc{r})}
   {E-E_{k_{\theta}}}
  \right],
  \nonumber \\
 \label{eq:lev_dens_CSM_2}
\end{eqnarray}
where $N_B$ and $N_R^\theta$ are the numbers of bound states and
resonances in the wedge region between the real energy axis and the
$2\theta$-lines, respectively.
A detailed explanation of the extended completeness relation is given in
Ref.~\citen{My98}.

In the integration over $\vc{r}$ in Eq.~(\ref{eq:lev_dens_CSM_2}), the
bound state and resonance parts are easily found to be unity, because of
the normalization of the wave functions, but the continuum part cannot
be calculated, due to the singular nature of the integration.
This singularity is eliminated when we discretize the continuum spectra
using the basis function method with a finite number $N$ of basis
functions. 
Then, the approximate density of states $\rho^{N}_{\theta}(E)$ for the
basis number $N$ is expressed as
\begin{eqnarray}
 \rho^{N}_{\theta}(E) 
  &=&
  \sum_{B}^{N_B}\delta(E-E_{B}) 
  - \frac{1}{\pi} {\rm Im} \sum_{R}^{N_{R}^{\theta}}\frac{1}{E-E_{R}}
  \nonumber\\
 & & \hspace{4cm}
  - \frac{1}{\pi} {\rm Im} \sum_{k}^{N-N_B-N_R^\theta}
  \frac{1}{E-{\cal E}_{k}(\theta)}.
  \label{eq:lev_dens_CSM_7}
\end{eqnarray}

As explained above, the energy of the resonance is obtained as
$E_{R}=E_{r}-i\Gamma_{r}/2$, and thus each resonance term has the
Breit-Wigner form
\begin{eqnarray}
 {\rm Im}\frac{1}{E-E_{R}}
  &=& 
  \frac{-\Gamma_{r}/2}{(E-E_{r})^2+\Gamma^{2}_{r}/4}.
  \label{eq:lev_dens_CSM_8}
\end{eqnarray}
For the continuum part, discretized continuum states are obtained on the
2$\theta$-line in the complex energy plane,
${\cal E}_k(\theta)={\cal E}_k^R-i{\cal E}_k^I$, where
 ${\cal E}_k^I/{\cal E}_k^R=\tan{2\theta}$.
Therefore, the continuum part in the level density can be expressed in
terms of a Lorentzian function whose form is similar to the Breit-Wigner
form:
\begin{eqnarray}
 {\rm Im}\frac{1}{E-{\cal E}_{k}(\theta)} 
  &=& 
  \frac{-{\cal E}_k^I}{(E-{\cal E}_{k}^R)^2+{{\cal E}_k^I}^{2}}.
  \label{eq:lev_dens_CSM_9}
\end{eqnarray}
Inserting Eqs.~(\ref{eq:lev_dens_CSM_8}) and (\ref{eq:lev_dens_CSM_9})
into Eq.~(\ref{eq:lev_dens_CSM_7}), we obtain the level density in the
basis function method as
\begin{eqnarray}
 \rho^{N}_{\theta}(E)
  &=& 
  \sum_{B}^{N_B}\delta(E-E_{B}) + \frac{1}{\pi} \sum_{R}^{N_{R}^{\theta}}
  \frac{\Gamma_{r}/2}{(E-E_{r})^2+\Gamma^{2}_{r}/4} 
  +\frac{1}{\pi} \sum_{k}^{N-N_B-N_R^\theta}
  \frac{{\cal E}_k^I}{(E-{\cal E}_{k}^R)^2+{{\cal E}_k^I}^{2}}.
  \nonumber\\ 
 \label{eq:lev_dens_CSM_10}
\end{eqnarray}

Here, it is noted that $\rho^{N}_{\theta}(E)$ has a $\theta$ dependence,
but $\rho(E)$ does not.
This $\theta$ dependence problem of $\rho^{N}_{\theta}(E)$ is due to the
fact that we employ a finite number of basis functions, and it can be
solved by introducing $\Delta(E)$ defined in Eq.~(\ref{delta1}).
The continuum level density (CLD) $\Delta(E)$ is expressed as a balance
between the density of states $\rho(E)$ obtained from the Hamiltonian
$H$ and the density of continuum states, $\rho_{0}(E)$,
obtained from the asymptotic Hamiltonian $H_0$ in the form
\begin{eqnarray}
 \Delta(E)  =  \bar{\rho}(E) -  \rho_{0}(E),
  \label{eq:CLD_Green}
\end{eqnarray}
where $\bar{\rho}(E)$ is defined through subtraction of the bound state
term from $\rho(E)$.
Physically, $\Delta(E)$ represents the density of unbound levels, which
result from the interaction with a finite range.
This can also be understood from the fact that $\Delta(E)$ is related to
the phase shift caused by the interaction.

In the basis function method with a finite number $N$ of basis states,
we have
\begin{eqnarray}
 \Delta^{N}_{\theta}(E)
  &=&
  \bar{\rho}_{\theta}^{N}(E)-\rho^{N}_{0(\theta)}(E).
  \label{eq:cld_csm_1}
\end{eqnarray}
The first term on the right-hand side represents the level density in
which the bound state term is subtracted from
Eq.~(\ref{eq:lev_dens_CSM_7}), and the second term is expressed in terms
of the eigenvalues
${\cal E}_k^0(\theta)={\cal E}_k^{0R}-i{\cal E}_k^{0I}$ of the
asymptotic Hamiltonian $H_0(\theta)$, which has only continuum spectra
on the $2\theta$-lines:
\begin{equation}
 \rho^{N}_{0(\theta)}(E)
  =
  \frac{1}{\pi} \sum_{k}^{N}
  \frac{{\cal E}_k^{0I}}{(E-{\cal E}_{k}^{0R})^2+{{\cal E}_k^{0I}}^{2}}.
\end{equation}
Thus, we have
\begin{eqnarray}
 \pi\Delta^{N}_{\theta}(E) 
  &=&
  \sum_{R}^{N_{R}^{\theta}}\frac{\Gamma_{r}/2}{(E-E_{r})^2+\Gamma^{2}_{r}/4} 
  +\sum_{k}^{N-N_B-N_R^\theta}
  \frac{{\cal E}_k^I}{(E-{\cal E}_{k}^R)^2+{{\cal E}_k^I}^{2}}
  \nonumber \\
 & &\hspace{5cm} -\sum_{k}^{N}
  \frac{{\cal E}_k^{0I}}{(E-{\cal E}_{k}^{0R})^2+{{\cal E}_k^{0I}}^{2}}. 
  \label{eq:cld_csm_2}
\end{eqnarray}
As shown by the numerical results presented in the next section, the
$\theta$ dependence of $\Delta^{N}_{\theta}(E)$ disappears through the
cancellation of the $\theta$ dependence in the second and third terms of
Eq.~(\ref{eq:cld_csm_2}).
When we consider a small value of $\theta$, and therefore no resonance
exists in the wedge region, the CLD can be expressed in terms of only
the second and third terms.

\section{Simple potential model}
We now examine the reliability of the present method for a simple
potential model.
As a schematic potential, we employ the CGKPM potential\cite{Cs90},
whose resonance structure has been studied in detail.
The Hamiltonian in this case is given by
\begin{eqnarray}
 H=T+V,\quad 
  T=-\frac{\hbar^{2}}{2\mu}\nabla^2,
  \quad
  V(r) = -8.0e^{-0.16r^{2}}+4.0e^{-0.04r^{2}},
  \label{eq:Hamiltonian}
\end{eqnarray}
where we set $\hbar^{2}/\mu$=1 (MeV/fm$^2$) for simplicity.
The Schr\"odinger equation for this Hamiltonian is solved by applying
the basis function method, and thus we write
\begin{equation}
 \psi(\vc{r})
  =
  \sum_{lm} R_\ell(r)Y_{\ell m}(\hat{r}),\quad
  R_{\ell}(r) =\sum_{i}^N c_{i}^{\ell}\phi_{\ell}(r,b_{i}).
\end{equation}
For each partial wave, we use Gaussian functions\cite{Ka88} with
different size parameters as basis functions:
\begin{equation}
 \phi_{\ell}(r,b_{i}) 
  =
  N_{\ell}(b_{i}) \cdot
  r^{\ell}\exp\left[-\frac{1}{2b_{i}^{2}}r^{2} \right]
  ,\quad
  N_{\ell}(b_{i})
  =
  b_{i}^{-3/2-\ell} 
  \left\{\frac{2^{\ell+2}}{(2\ell+1)!!\sqrt{\pi}} \right\}^{1/2}, 
  \label{eq:GaussBase}
\end{equation}
where the parameters $\{b_{i}: i=1,2,\cdots,N \}$ are given by a
geometrical progression\cite{Ka88} of the form
\begin{eqnarray}
 b_{i}&=&b_{0}\gamma^{i-1}.
\end{eqnarray}
Here, $b_{0}$ and $\gamma$ are the first term and the common ratio,
respectively.
We employ $N=30$, $b_{0}=0.2$ fm and $\gamma=1.2$ in the following
calculations.
Of course, the same results are obtained even if other kinds of basis
functions (for example, harmonic oscillator functions) are used.
\begin{figure}[tbp]
 \begin{minipage}[b]{\halftext}
  \centerline{\includegraphics[width=6.6cm]{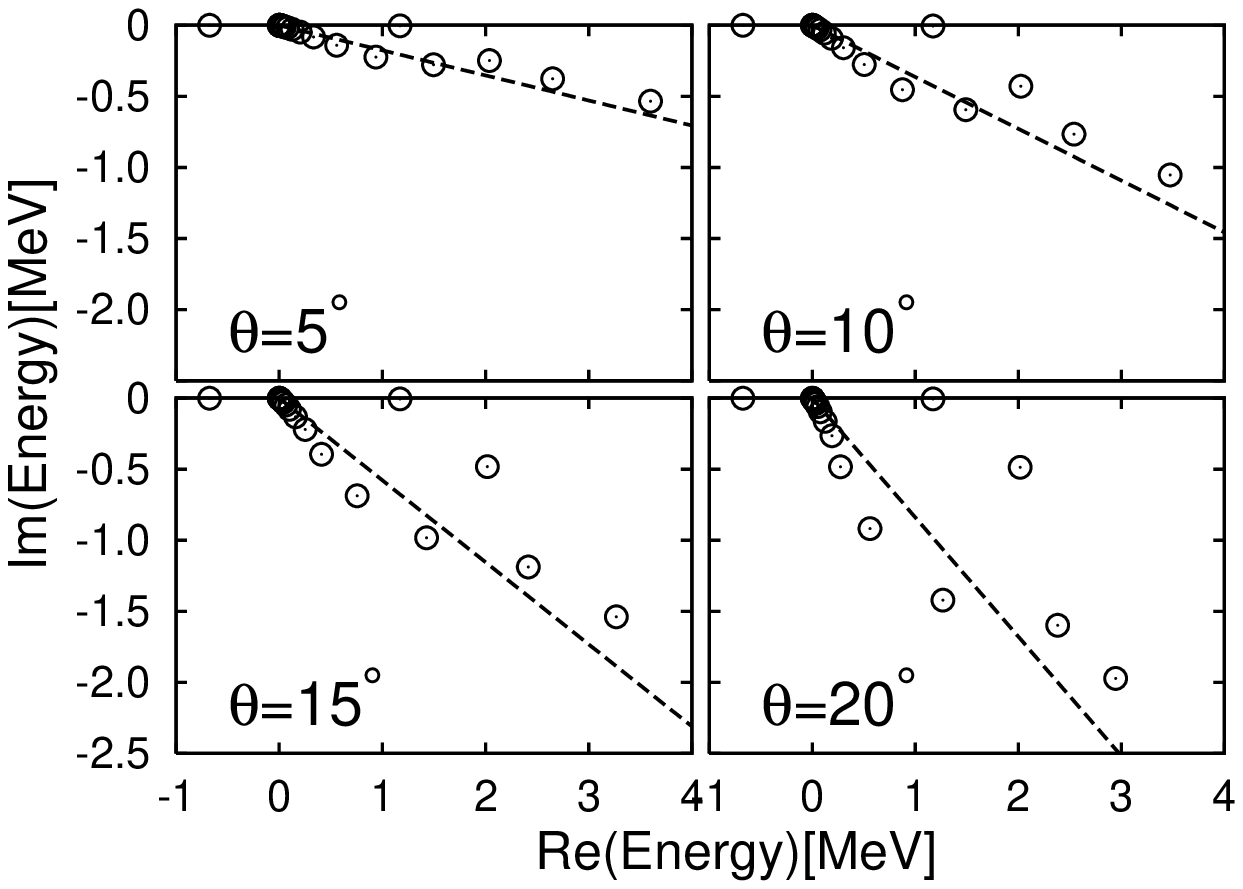}}
  \caption{
  Energy eigenvalue distribution of the $1^-$ states for the
  complex scaled Hamiltonian of the simple potential model given by
  Eq.~(\ref{eq:Hamiltonian}).
  The circles represent eigenvalues and dashed lines are
  2$\theta$-lines.
  }
  \label{fig:Energy_N30}
 \end{minipage}
 \hspace{0.6cm}
 \begin{minipage}[b]{\halftext}
  \centerline{\hspace{2.2cm}\includegraphics[width=8.8cm]{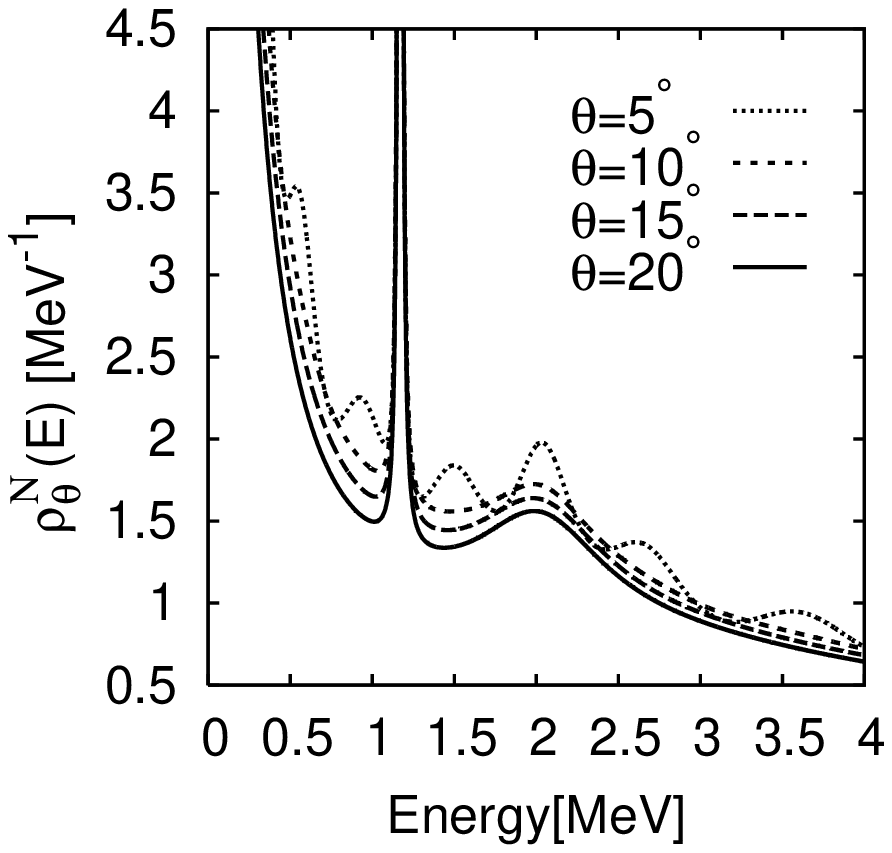}}
  \caption{
  The level density $\rho^{N}_{\theta}(E)$ calculated for different
  values of $\theta$.
  }
  \label{fig:density_of_state_N30}
 \end{minipage}
\end{figure}

In Fig.~\ref{fig:Energy_N30}, we plot the energy eigenvalue distribution
of the 1$^-$ states.
One bound state ($-0.67$ MeV) and many resonant states exist: The three
lowest resonances are $1.1710-i0.0049$ (MeV), $2.0175-i0.4863$ (MeV) and
$2.5588-i1.7378$ (MeV).\cite{Ho97}
The lowest resonance is obtained with the CSM for $\theta=5^\circ$, but
the second lowest resonance is not obtained for this scaling parameter
value.
The second resonance appears clearly when $\theta>10^\circ$.
The continuum solutions vary slightly from the $2\theta$-line, and the
dispersion increases for large values of $\theta$.
However, this distribution of continuum eigenvalues depends on the
choice of the basis functions, and it creates no difficulty in the CLD
calculations.

Using these eigenvalues, we calculate the level density
$\rho^{N}_{\theta}(E)$ given by Eq.~(\ref{eq:lev_dens_CSM_7}), and we
display the result in Fig.~\ref{fig:density_of_state_N30}.
Oscillatory behavior is seen at $\theta=5^\circ$, but this oscillation
is smoothed when $\theta$ is larger than $10^\circ$.
Even at $\theta=5^\circ$, the oscillation may disappear if we employ a
large number of basis functions so that the intervals between the
discretized continuum eigenvalues become smaller than their imaginary
parts.
However, it is easier to choose a larger value of $\theta$ in order to
increase the imaginary parts of the discretized continuum eigenvalues.
The intervals between the discretized continuum eigenvalues depend on
the number $N$ of basis functions.
The critical value of $\theta$ may be defined as the scaling angle at
which the imaginary parts of the discretized continuum eigenvalues
become larger than the intervals between the eigenvalues.
This critical value of $\theta$ depends on $N$, and therefore we express
it as $\theta_N$.
When $\theta$ becomes larger than $10^\circ$ in the present simple
potential case, $\rho^{N}_{\theta}(E)$ exhibits the same behavior, and
therefore we can set $\theta_N\approx 10^\circ$.
For $\theta>\theta_N$, only the absolute values of
$\rho^{N}_{\theta}(E)$ depend on $\theta$.

\begin{figure}[tbp]
 \centerline{\hspace{2.0cm}\includegraphics[width=11.3cm]{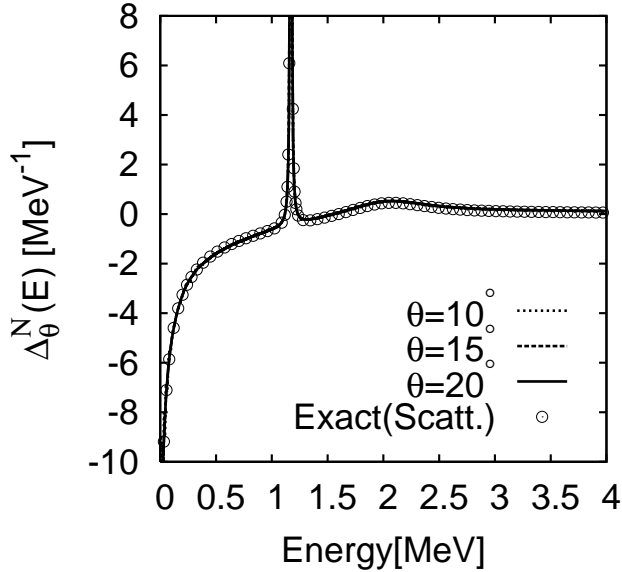}}
 \caption{
 The complex scaled CLD $\Delta^N_\theta(E)$ for
 $\theta=10^\circ,~15^\circ$ and $20^\circ$ and the exact solution
 obtained by solving the scattering problem.
 }  
 \label{fig:CSCLD_comp}
\end{figure}

This $\theta$ dependence of the absolute values of $\rho^N_\theta(E)$
can be canceled through subtraction of $\rho^{N}_{0\theta}(E)$; that is,
we show that the CLD $\triangle^{N}_{\theta}(E)$ defined in
Eq.~(\ref{eq:cld_csm_2}) has no $\theta$ dependence for
$\theta\ge\theta_N$.
In Fig.~\ref{fig:CSCLD_comp}, we plot the CLD
$\triangle^{N}_{\theta}(E)$ calculated for $\theta=10^\circ,\ 15^\circ$
and $20^\circ$ and compare it with the result of the exact calculation.
Here, ``exact'' means that we calculate the CLD $\triangle(E)$ from the
phase shift using Eq.~(\ref{eq:exact}).
The phase shift is obtained with the help of the scattering solution
without any approximation.
From Fig.~\ref{fig:CSCLD_comp}, we see that it is quite difficult to
distinguish the plots of $\triangle^{N}_{\theta}(E)$ calculated for
$\theta=10^\circ,\ 15^\circ$ and $20^\circ$.
They are all consistent with the exact calculation.
This result indicates that the CLD $\triangle(E)$ can be approximated by
$\triangle^{N}_{\theta}(E)$ in the CSM, and the phase shift can be
obtained from $\triangle^{N}_{\theta}(E)$ without solving the scattering
problem. 

\section{Applications to $^4$He+$n$ and $\alpha+\alpha$ systems}
We now apply the present method to realistic two-body systems of
$^4$He+$n$ and $\alpha+\alpha$. 
The $^4$He+$n$ system has rather broad resonances but no bound state,
and the $\alpha+\alpha$ system also has no bound state but a sharp
resonance due to the Coulomb barrier.
The Coulomb potential is a typical long-range potential and is
represented by the asymptotic term of $H_0$.
The antisymmetrization among clusters in both the systems $^4$He+$n$ and
$\alpha+\alpha$ is carried out with the orthogonality condition model
(OCM).\cite{Sa69}
We show that the present method is very useful in analyses of continuum
states of such realistic cluster systems.
  
\subsection{$^4$He+$n$ system}
The wave function of $^5$He with spin $J$ is expressed in the $^4$He+$n$
cluster model as
\begin{equation}
 \Phi^J(^5{\rm He})
  =
  {\cal A}\left\{\Phi(^4{\rm He})\cdot \psi^J_{\rm rel}(\vc{r})\right\},
\end{equation}
where ${\cal A}$, $\Phi(^4{\rm He})$ and $\psi^J_{\rm rel}(\vc{r})$ are
the antisymmetrizer, the internal wave function of $^4$He assuming a
$(0s_{1/2})^4$ configuration, and the relative wave function between
$^4$He and the valence neutron, respectively.
We solve the relative wave function $\psi^J_{\rm rel}(\vc{r})$ by
applying the OCM.
This yields
\begin{equation}
 \left[
  T_{\rm rel} + V_{\alpha n}(r) 
  + \lambda\, |\phi_{\rm PF}\rangle\langle \phi_{\rm PF}| -E 
 \right]
\psi^J_{\rm rel}(\vc{r})=0,
\label{5HeOCM}
\end{equation}
where $T_{\rm rel}$ and $V_{\alpha n}(r)$ are the kinetic energy and
potential operators for the $^4$He-$n$ relative motion, respectively.
In this calculation, we use the so-called KKNN potential\cite{Ka79} for
$V_{\alpha n}(r)$, which provides an accurate description of the
low-energy scattering data for this system.
The third term, constituting the non-local potential in
Eq.~(\ref{5HeOCM}) represents the projection operator to remove the
Pauli forbidden (PF) states [which is the $(0s_{1/2})$ state in this
case] from the relative motion\cite{Ku86}, and $\lambda$ is taken as
$10^6$~MeV in this calculation.
\begin{figure}[b]
 \centerline{\includegraphics[width=12.3cm]{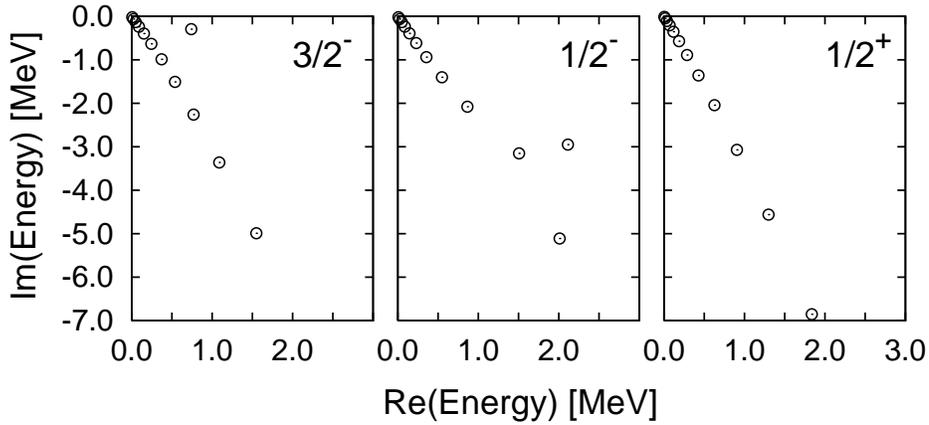}}
 \caption{
 Energy eigenvalue distributions of the $^4$He-$n$ system for
 the $J^\pi=3/2^-,1/2^-$ and $1/2^+$ states, where $\theta$ is taken as
 35$^\circ$.
 } 
 \label{fig:5He_ene}
\end{figure}
\begin{figure}[t]
 \centerline{\includegraphics[width=13.3cm]{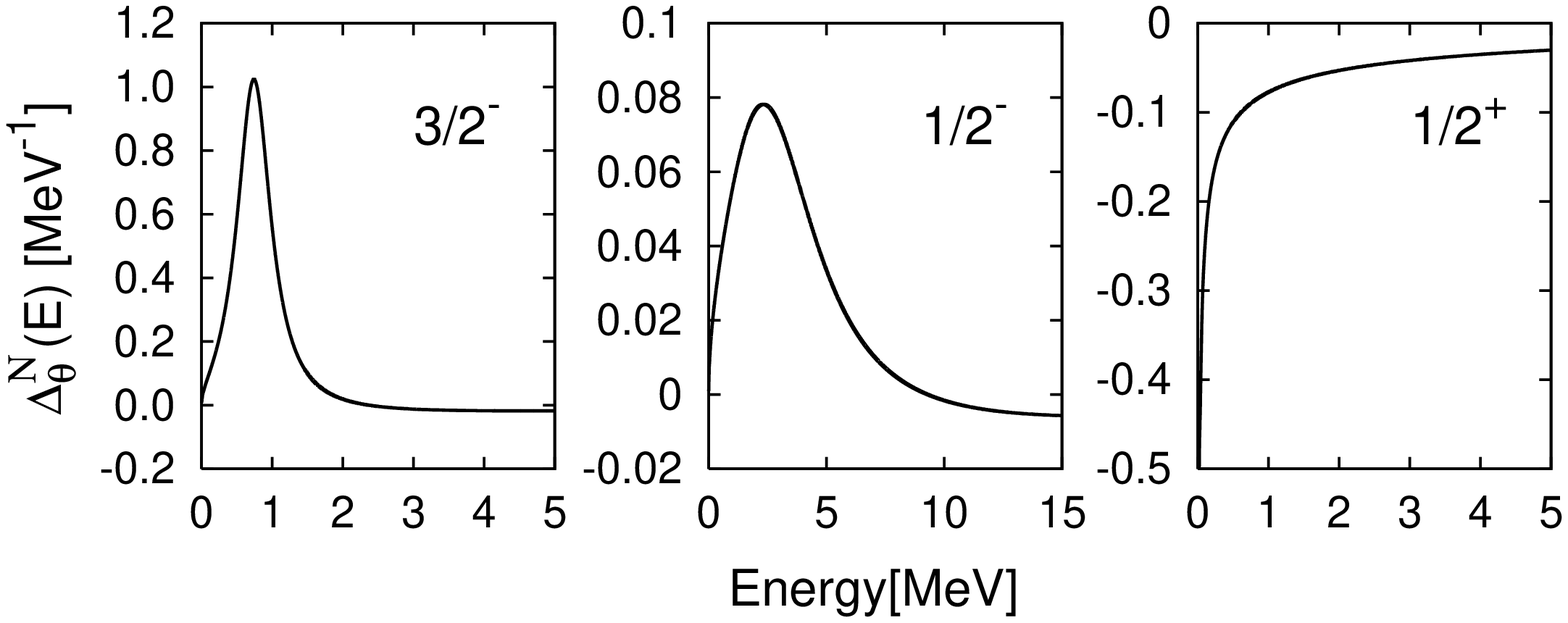}}
 \caption{
 Continuum level densities of the $^4$He-$n$ system for the
 $J^\pi=3/2^-,1/2^-$ and $1/2^+$ states.
 }
 \label{fig:5He_ld}
\end{figure}

Equation (\ref{5HeOCM}) is solved by using the basis functions, as
explained in the previous section, and we obtain 
\begin{equation}
 \psi^J_{\rm rel}(\vc{r})
  =
  \left[Y_\ell (\hat{r})\chi_{1/2}\right]_J\varphi_\ell(r),\hspace{1cm}
  \varphi_\ell(r)=\sum_i^Nc^\ell_i\phi_\ell(r,b_i),
\end{equation}
where $\left[Y_\ell (\hat{r})\chi_{1/2}\right]_J$ is a function of the
orbital angular momentum and spin coupled to $J$, and the radial wave
function $\varphi_\ell(r)$ is expanded in the Gaussian basis functions
$\{\phi_\ell(r,b_i)\}$, which are defined in Eq.~(\ref{eq:GaussBase}).
\begin{figure}[bp]
 \centerline{\includegraphics[width=10.5cm]{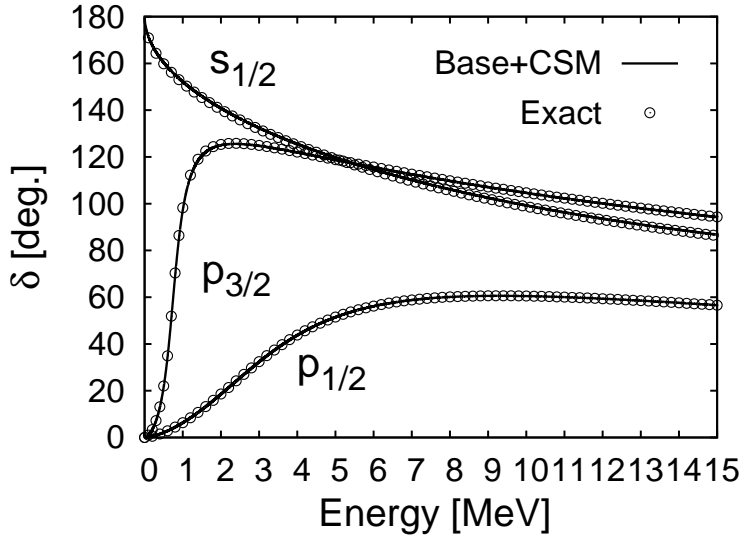}}
 \caption{
 Scattering phase shifts of the $^4$He-$n$ system for the
 $J^\pi=3/2^-,1/2^-$ and $1/2^+$ states.
 }
 \label{fig:5He_ph}
\end{figure}

Using the same basis set as in the case of the simple model, we calculate
the energy eigenvalues of the complex scaled Hamiltonian with
$\theta=35^\circ$, and the results for the three states $3/2^-$, $1/2^-$
and $1/2^+$ are shown in Fig.~\ref{fig:5He_ene}.
We can see that each of the states $3/2^-$ and $1/2^-$ has one resonance
pole, corresponding to the observed resonances of $^5$He.
The $1/2^+$ state has no resonance.
Resonant structures of $^5$He have been investigated in detail with the
complex scaling method by Aoyama et al.\cite{Ao95}
In addition to resonances, the discretized continuum solutions have been
obtained along the $2\theta$-line.
Several continuum solutions are off the $2\theta$-line.
It is believed that the reason for this is that the couplings between
the continuum and resonance are not correctly described because the
number of basis functions is not large enough.
However, the resonant solutions are obtained with appropriate accuracy,
and the CLD is obtained from these continuum solutions satisfactorily,
although the positions of some continuum solutions are slightly off the
$2\theta$-line.   

Applying Eq.~(\ref{eq:cld_csm_2}) to the obtained eigenvalue
distribution of the complex scaled Hamiltonian for the $3/2^-$, $1/2^-$
and $1/2^+$ states, we calculate the CLD of the $^4$He-$n$ system.
The results are shown in Fig.~\ref{fig:5He_ld}.
It is seen that each of the $3/2^-$ and $1/2^-$ states has a peak, but
the $1/2^+$ state has no peak.
The position and width of the peaks in the CLD for the $3/2^-$ and
$1/2^-$ states corresponds to their resonance energy and width.
These results are very similar to those for the CLD distributions
calculated by Arai and Kruppa,\cite{Ar99} except for the absolute
strengths.
In the calculation of the CLD carried out by Arai and Kruppa, the result
depends on the smoothing parameter.  
  
To see the reliability of the CLD obtained here, we calculate the phase
shift from the obtained CLD.
In Fig.~\ref{fig:5He_ph}, we show the scattering phase shifts of the
$3/2^-$, $1/2^-$ and $1/2^+$ states.
We compare these results with the exact phase shifts, and we find very
good quantitative agreement between them for every state. 
\subsection{$\alpha+\alpha$ system}
Similarly to the above, we now calculate the CLD and the scattering
phase shifts of the $\alpha+\alpha$ system.
The important point in the calculation of the $\alpha+\alpha$ system is
the treatment of the Coulomb interaction.
Because the Coulomb interaction has a long-range nature, we must include
the Coulomb interaction in the asymptotic Hamiltonian $H_0$.
\begin{figure}[b]
 \centerline{\includegraphics[width=13.3cm]{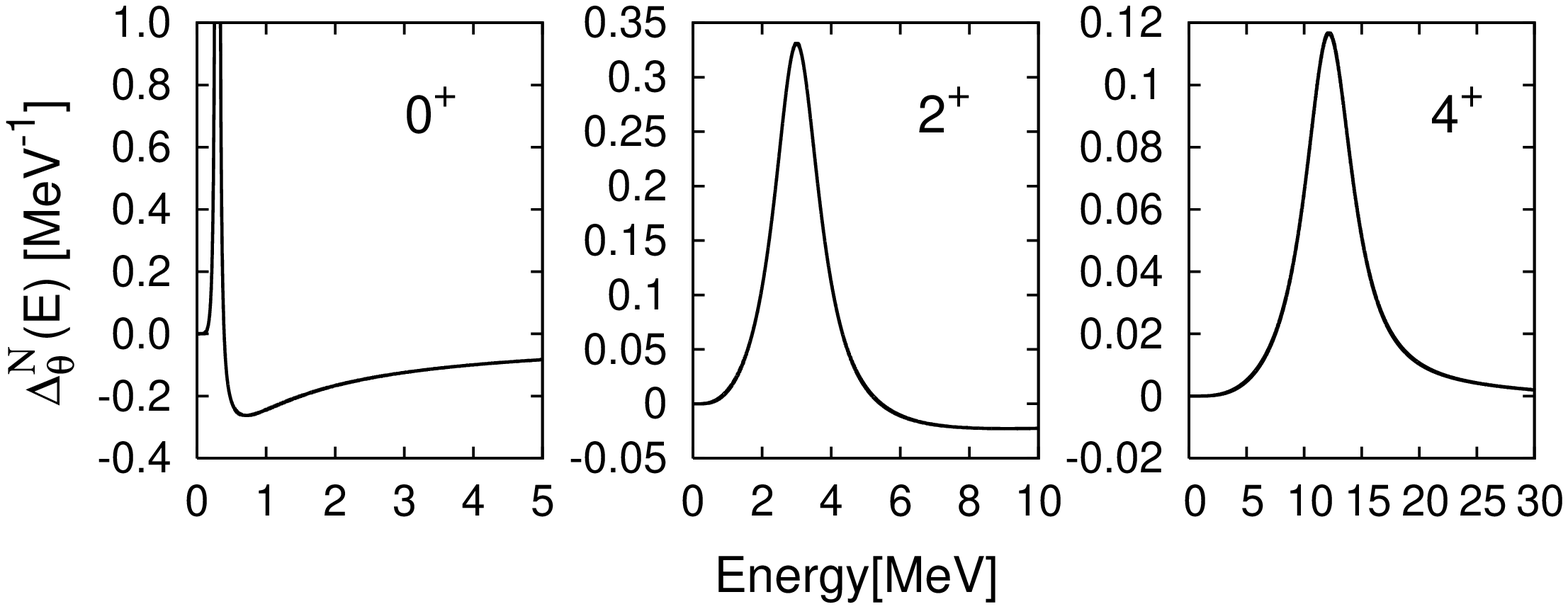}}
 \caption{
 Continuum level densities of the $\alpha$-$\alpha$ system for
 the $J^\pi=0^+,\ 2^+$ and $\ 4^+$ states.
 }
 \label{fig:8Be-CLD}
\end{figure}
  
The relative motion between two $\alpha$ clusters is described within
the OCM as
\begin{equation}
 \left[
  T_{\rm rel}+V_{\alpha\alpha}^C(r)+V_{\alpha\alpha}^N(r)
  +\lambda\sum_{\rm PF}|\phi_{\rm PF}\rangle\langle \phi_{\rm PF}| -E 
 \right]
 \psi^J_{\rm rel}(\vc{r})=0,
 \label{2alphaOCM}
\end{equation}
where $V_{\alpha\alpha}^C$ and $V_{\alpha\alpha}^N$ are the folding
Coulomb and nuclear potentials obtained by assuming a $(0s_{1/2})^4$
harmonic oscillator wave function with oscillator constant
$\nu_{\alpha}(=\frac{M\omega}{2\hbar})=0.2675$ fm$^{-2}$ for an $\alpha$
cluster, respectively.
When we employ the Schmid-Wildermuth force\cite{Sh61} as the two-nucleon
force, they are expressed as
\begin{eqnarray}
 V_{\alpha\alpha}^C(r)
  &=&
  \left(\frac{4e^2}{r}\right){\mbox e}{\mbox r}{\mbox f}
  \left(r\sqrt{\frac{4}{3}\nu_{\alpha}}\right),\\
 V_{\alpha\alpha}^N(r)
  &=&
  2X_D \left[ \frac{2\nu_{\alpha}}{2\nu_{\alpha}+3\mu/2} \right]^{3/2}
  V_{0}\exp\left[-\frac{\nu_{\alpha}\mu}{\nu_{\alpha}+3\mu/4}r^2\right],
\end{eqnarray}
where erf$(x)$ is the error function, and $X_D=2.445$, $V_0=-72.98$ MeV
and $\mu=0.46$ fm$^{-2}$ are the folding parameter, the strength, and
the range parameter of the Schmid-Wildermuth force, respectively.
The fourth term in Eq.~(\ref{2alphaOCM}) is the projection operator to
remove the Pauli forbidden states (the $0S,~1S$ and $0D$ states in this
case) from the relative motion,\cite{Ku86} and $\lambda$ is taken as
$10^6$ MeV as well.
We solve the complex scaled Schr\"odinger equation Eq.~(\ref{2alphaOCM})
in the same way as we solved the Schr\"odinger equation for the simple
potential and $^4$He-$n$ systems.
Using the obtained eigenvalues for $J^\pi=0^+,~2^+$ and $4^+$, we
calculate the CLD.
In the $\alpha$-$\alpha$ system, however, the eigenvalues of the
asymptotic Hamiltonian $H_0$ must be solved with the Coulomb potential:
\begin{equation}
 H_{0}= T_{\rm rel} + \frac{4e^2}{r}.
\end{equation}
The results of the CLD are shown in Fig.~\ref{fig:8Be-CLD}.
They have a sharp peak corresponding to the resonance in each state.
This result is quite similar to the results
by Arai and Kruppa\cite{Ar99}, in which case the smoothing was performed
in terms of several smoothing parameters.
Better agreement is obtained for a narrower smoothing parameter.

\begin{figure}[tbp]
 \centerline{\includegraphics[width=10.6cm]{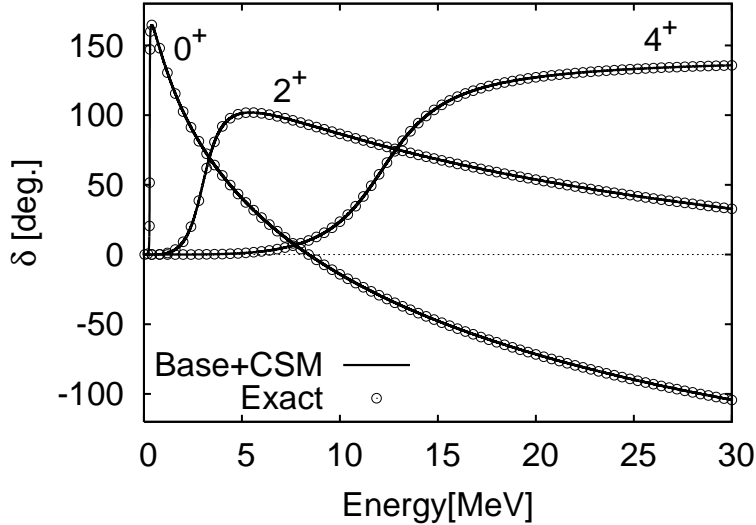}}
 \caption{
 Scattering phase shifts of the $\alpha$-$\alpha$ system for the
 $J^\pi=0^+,\ 2^+$ and $\ 4^+$ states.
 }
 \label{fig:8Be-phase}
\end{figure}
Integrating the obtained CLD, we obtain the scattering phase shifts.
The results are shown in Fig.~\ref{fig:8Be-phase}.
The scattering phase shifts are nearly identical to those obtained from
the scattering solutions.
The resonance width of the $0^+$ state is very small in comparison to
the resonance energy.
For such a case, it is necessary to carefully integrate the CLD to
obtain accurate phase shifts.
These results indicate that the present method to calculate the CLD is
also very powerful even for a long-range interaction, such as the
Coulomb potential.
%

\section{Summary and conclusion}
We have shown that the level density is properly described in the CSM
with a basis function method.
In the expression we obtained for the level density, the extended
completeness relation of the CSM plays an important role, and it divides
the level density into three terms, i.e., bound states, resonances and
continuum states.
We investigated the approximate description of the continuum states in
terms of discretized eigenstates that are obtained through
diagonalization with a finite number of basis functions.
Furthermore, it is not necessary to use a smoothing technique, such as
the Strutinsky procedure employed by Kruppa and Arai for the singular
level density arising from the discretization of continuum states.
In the CSM, continuum states are expressed in terms of eigenstates of
complex eigenvalues along the rotated branch cut with the angle
$2\theta$, and the Green function of the continuum part is expressed as
a sum of Lorentzian functions rather than delta functions.
Therefore, no singularity appears.
This result indicates that the CSM provides a very powerful method for
discretizing continuum states.
The discretization of a continuous function using Lorentzian functions
would be understood through comparison to the wavelets\cite{Da92} that
have recently been developed as a powerful tool facilitating
transformations between analogue and digital data in information
science. 

The level density smoothed in the CSM, however, has a dependence on the
scaling angle $\theta$, because a finite number of basis functions is
used in the approximate description of the continuum states.
We showed that the continuum level density (CLD) in the CSM, obtaining
by subtracting the level density for the asymptotic Hamiltonian, is
independent of the scaling angle and consistent with the exact CLD.
These results indicate that we can calculate scattering phase shifts or
S-matrices from the CLD obtained by solving an eigenvalue problem in the
CSM with a finite number of basis functions.
We found that this method is quite effective in the treatment of
$^4$He+$n$ and $\alpha+\alpha$ systems without and with the Coulomb
interaction, respectively, which were previously studied by Arai and
Kruppa.\cite{Ar99}

Considering the successful results of this method for simple two-body
systems, it would be interesting to apply it to coupled-channel systems
and three-body systems.
For coupled-channel problems, the extended completeness relation
providing the foundation of this method has been proven in the framework
of the CSM.\cite{Gi04}
Therefore, it is conjectured that the present method will be effective
here too.
However, the three-body problem is still open.

\section*{Acknowledgements}
The authors would like to thank Dr. K. Arai for fruitful discussions. 
They also would like to acknowledge the members of the nuclear theory
group at Hokkaido University for many discussions.
This work was performed as a part of the ``Research Project for Study of
Unstable Nuclei from Nuclear Cluster Aspects (SUNNCA)'' sponsored by
RIKEN.


\end{document}